\documentclass[11pt]{article}

\usepackage{multirow}

\usepackage[nottoc]{tocbibind}

\usepackage{geometry}
\usepackage[affil-it, auth-sc]{authblk}
\usepackage[utf8]{inputenc}

\usepackage{caption}
\usepackage{subcaption}

\usepackage{booktabs}

\usepackage{graphicx}
\usepackage{amssymb}
\usepackage{amsthm}
\usepackage{amsmath}
\usepackage{color, soul}
\usepackage{verbatim}
\usepackage{lineno}
\usepackage{todonotes}
\presetkeys{todonotes}{inline, color=blue!30, size=\small}{}

\definecolor{darkred}{rgb}{0.9, 0.0, 0.0}
\definecolor{darkgreen}{rgb}{0.0, 0.5, 0.0}
\usepackage[pdftex,colorlinks,bookmarks,linkcolor=darkred, citecolor=darkgreen]{hyperref}

\usepackage[ampersand]{easylist}

\usepackage{bm}
\usepackage{feynmf}

\newcommand{\be}{\begin{equation}}
\newcommand{\ee}{\end{equation}}
\def\ba{\begin{eqnarray}}
\def\ea{\end{eqnarray}}

\usepackage[sort&compress,numbers]{natbib}
\usepackage{doi}

\usepackage{slashed}

\newcommand{\nl}{\nonumber \\ }

\makeatletter
\newcommand{\vast}{\bBigg@{4}}
\newcommand{\Vast}{\bBigg@{5}}
\makeatother

\usepackage{eso-pic}

\allowdisplaybreaks

\begin{document}

\AddToShipoutPictureFG*{

    \AtPageUpperLeft{\put(-60,-75){\makebox[\paperwidth][r]{FERMILAB-PUB-19-564-T}}}  
    }

\title{\bf Direct detection rate of heavy Higgsino-like and Wino-like dark matter}

\date{\today}

\author[1]{Qing Chen}
\author[1,2]{Richard J. Hill}
\affil[1]{Department of Physics and Astronomy, University of Kentucky, Lexington, KY 40506, USA \vspace{1.2mm}}
\affil[2]{Theoretical Physics Department, Fermilab, Batavia, IL 60510, USA \vspace{1.2mm}}

\maketitle

\begin{abstract}
Many viable dark matter models contain a WIMP candidate that is
a component of a new electroweak multiplet whose mass $M$ is large compared to the electroweak scale $m_W$.  A generic amplitude-level cancellation in such models yields a severe suppression of the cross section for WIMP-nucleon scattering,   
making it important to assess the impact of formally subleading effects.  The power correction of order $m_W/M$ to the heavy WIMP limit is computed for electroweak doublet (Higgsino-like) dark matter candidates, and a modern model of nuclear modifications to the free nucleon cross section is evaluated.  
Corrections to the pure Higgsino limit are determined by a single parameter through first order in the heavy WIMP expansion.  Current and projected experimental bounds on this parameter are investigated.  The direct detection signal in the pure Higgsino limit remains below neutrino backgrounds for WIMPs in the TeV mass range.  Nuclear corrections are applied also to the heavy Wino case, completing the investigation of combined subleading effects from perturbative QCD, $1/M$ power corrections, and nuclear modifications.  
\end{abstract}

\newpage

\section{Introduction} 

The Weakly Interacting Massive Particle (WIMP) has long been considered as a well-motivated dark matter candidate~\cite{Lee:1977ua,Gunn:1978gr,Steigman:1997vs,Ellis:1983ew,Goodman:1984dc,Goldberg:1983nd,Bertone:2004pz, Feng:2010gw,Bramante:2015una,Arcadi:2017kky,Roszkowski:2017nbc} but remains experimentally undetected~\cite{Tanabashi:2018oca}.  WIMPs naturally fit in the paradigm of supersymmetric extensions of the Standard Model~\cite{Jungman:1995df} yet we haven't found evidence at the LHC for supersymmetric particles at the electroweak scale~\cite{ATLAS-CONF-2019-014, Longo:2687347}.  Within the WIMP paradigm, the situation is suggestive that new particles are somewhat heavy compared to the electroweak scale, and in particular, $M_{\rm WIMP} \gg M_{W^\pm}, M_{Z^0}$.  In this mass regime, heavy WIMP effective field theory becomes a powerful method to study the universal behavior in low energy WIMP-nucleus scattering processes~\cite{Hill:2011be, Hill:2014yka, Hill:2014yxa},  predicting cross sections for dark matter direct detection experiments that are minimally sensitive to unknown ultraviolet (UV) physics.  

Explicit calculations in heavy WIMP effective theory~\cite{Hill:2013hoa,Hill:2014yka, Hill:2014yxa} reveal an amplitude level cancellation~\cite{Hill:2011be,Hill:2013hoa,Hisano:2011cs} that results in cross section predictions for electroweak triplet (Wino-like) and electroweak doublet (Higgsino-like) WIMPs that are below the sensitivity of current direct detection experiments~\cite{Aprile:2018dbl}.  Such particles thus remain as viable dark matter candidates but it is important to understand whether naively subleading effects could alter the predicted cross section and hence their experimental observability. 

To improve the leading order calculation and to compare with next-generation experiments~\cite{ Aprile:2015uzo, Akerib:2015cja,Amaudruz:2017ekt, Liu:2017drf} approaching the neutrino floor~\cite{Billard:2013qya}, we consider subleading effects from the following sources.  First, $1/M$ power corrections in the heavy WIMP expansion depend on the specific representation of electroweak ${\rm SU}(2)\times {\rm U}(1)$ symmetry, and on the detailed UV completion of the WIMP theory.  For the case of electroweak triplet, power corrections for the pure Wino case were themselves found to exhibit a surprising level of cancellation~\cite{Chen:2018uqz}, yielding a cross section prediction for low velocity WIMP-nucleon scattering of $\sigma \sim 10^{-47}\,{\rm cm}^2$, for $M \gtrsim 500\,{\rm GeV}$.  Given that Higgsino-nucleon scattering suffers an even more severe amplitude cancellation compared to the Wino case~\cite{Hill:2014yxa}, it is important to study the power corrections in this case.  We also explore the consequences of structure beyond the pure Higgsino limit. 
Second, a complete accounting of nuclear effects can potentially alter the predicted direct detection event rate compared to simple models that apply a nuclear form factor to the single nucleon cross section. 
Since the cancellation occurs between single nucleon matrix elements of scalar and tensor currents, nuclear effects could be effectively enhanced by impacting the scalar and tensor currents differently.  We estimate the impact of such nuclear effects for both triplet and doublet cases, employing a recent model that incorporates constraints of chiral symmetry and multibody interactions~\cite{Hoferichter:2018acd, Hoferichter:2016nvd}.

The rest of this paper is organized as follows.  Section~\ref{sec:higgsino} constructs the heavy WIMP effective theory for Higgsino-like particles through $1/M$ order.  Section~\ref{sec:matching} performs electroweak scale matching for the pure Higgsino onto heavy WIMP effective theory, and specifies the treatment of renormalization group evolution and heavy quark threshold matching.  Section~\ref{sec:results} provides cross section results and discusses the impact of $1/M$ corrections for Higgsino-like. Section~\ref{sec:nuclear} considers nuclear modifications for the direct detection rate of Higgsino-like and Wino-like particles. Section~\ref{sec:summary} is a summary.

\section{Heavy Higgsino effective theory at order {\boldmath $1/M$} \label{sec:higgsino}}

Heavy WIMPs with mass $M$ large compared to the electroweak scale may be described using an effective theory expanded in powers of $1/M$.  Each order is constructed from invariant operators built from Standard Model fields and the heavy WIMP field; the latter transforms as an ${\rm SU}(2)_W\times {\rm U}(1)_Y$ multiplet and is denoted by $\chi_v$.  For a heavy self conjugate (e.g. Majorana fermion) particle the Heavy WIMP Effective Theory Lagrangian up to $1/M$ order takes the following form in the one-heavy particle sector (cf. Refs.~\cite{Hill:2011be, Chen:2018uqz}):
\ba
\mathcal{L}_{\rm HWET}=
\bar{\chi}_v\bigg[i v\cdot D- \delta M-\frac{{D}_\perp^2}{2M}-
\frac{f(H)}{M}-\frac{g(W,B)}{M}+ \dots \bigg]\chi_v \,,
\label{EFT}
\ea
where $v^\mu$ is the heavy WIMP velocity with $v^2=1$.  The covariant derivative is $D_\mu=\partial_\mu- i g_1 Y B_\mu-i g_2 W_\mu^a t^a$ and $D_\perp^\mu = D^\mu - v^\mu v\cdot  D$.  Dimension five operators $\bar{\chi}_vf(H)\chi_v$ and $\bar{\chi}_v g(W,B)\chi_v$ describe WIMP interactions with the Higgs field $H$ and with the electroweak field strengths $W_{\mu\nu}$ and $B_{\mu\nu}$, respectively. $\delta M$ is the residual mass.
In the following, we focus on the spin-independent process and neglect the field strength interaction term  $g(W,B)$, which only contributes to spin-dependent scattering.%
\footnote{Although the spin-independent amplitude suffers a severe cancellation at leading order in $1/M$, the spin-dependent amplitude vanishes at leading order.  Since it lacks the coherent enhancement of the spin-independent amplitude, this contribution is expected to remain numerically subdominant in the total direct detection rate.}

Let us consider a Standard Model extension whose particle content consists of a Dirac fermion WIMP transforming as an SU$(2)$ doublet with hypercharge $Y=1/2$.  This situation may arise in
models with supersymmetry~\cite{Jungman:1995df,Nagata:2014wma,Krall:2017xij} and extra dimensions~\cite{Delgado:2018qxq}.  Related models involve scalars~\cite{Cohen:2011ec,Chao:2018xwz}. 
We anticipate the splitting of mass eigenstates into Majorana components after electroweak symmetry breaking, and write the gauge invariant WIMP-Higgs interaction term $f(H)$ in the Majorana basis as
\ba
f(H)
=\begin{pmatrix}
a\mathbf{Re}(HH^\dagger)+\mathbf{Re}(b HH^T)+cH^\dagger H &
a\mathbf{Im}(HH^\dagger)-\mathbf{Im}(b HH^T)\\
-a\mathbf{Im}(HH^\dagger)-\mathbf{Im}(b HH^T)
& a\mathbf{Re}(HH^\dagger)-\mathbf{Re}(b HH^T)+cH^\dagger H 
\end{pmatrix} \,.
\label{int}
\ea
Here  the real parameters $a$ and $c$, and the complex parameter $b$,
are determined by matching with a specific UV theory.

To investigate the impact of additional UV structure, we consider a 
simple illustration where, in addition to the Dirac doublet $\psi$ of mass $M$, the Standard Model extension includes another ${\rm SU}(2)$ multiplet with a mass greater than $M$~\cite{Hill:2013hoa,Berlin:2015njh}.
For example, consider an ${\rm SU}(2)$ triplet Majorana fermion $\chi^\prime$ with mass $M^\prime\gg M$ (another interesting case is a heavy singlet).
The renormalizable Lagrangian is
\ba
\mathcal{L}_{\rm UV}=\mathcal{L}_{\rm SM}+
\bar{\psi}(i\slashed{D}-M)\psi
+\frac{1}{2}\bar{\chi}^\prime (i\slashed{D}-M^\prime)\chi^\prime
-\frac{1}{2}\bar{\lambda} F(H) \lambda \,,
\label{UV}
\ea
where 
$\lambda=
\begin{pmatrix}
\chi^\prime,& \chi_1, & \chi_2
\end{pmatrix}^T
$, with $\chi_1 = (\psi + \psi^c)/\sqrt{2}$ and $\chi_2 = i(\psi-\psi^c)/\sqrt{2}$. 
Note that $\chi=\begin{pmatrix} \chi_1, & \chi_2 \end{pmatrix}^T$
is the relativistic field mapping onto the heavy particle field $\chi_v$ 
in Eq.~(\ref{EFT}). 
$F(H)$ is the interaction with the Higgs field and can be found in Ref.~\cite{Hill:2014yka}.
Introduction of the heavy multiplet, $\chi^\prime$, splits the mass of the two neutral constituents $h_0^{\rm high}$, $h_0^{\rm low}$ of $\chi_v$ after electroweak symmetry breaking, with the lighter state $h_0^{\rm low}$ being identified as the dark matter WIMP.   By appropriate field redefinition, the residual mass of $h_0^{\rm low}$ may be set to zero.
Including the electrically charged eigenstates $h_+$ and $h_-$, 
the residual mass matrix in the mass-electric charge eigenstate basis $\big(h_0^{\rm high},h_0^{\rm low},h_+, h_- \big)$ 
is
 \ba
\delta M + \frac{f(\langle H \rangle )}{M} = \frac{v^2}{2M} {\rm diag}\begin{pmatrix}2|b|,&0,& |b|-a,&  |b|-a\end{pmatrix} \,,
\label{eq:residual}
\ea
where $\langle H \rangle = (0,v)^T/\sqrt{2}$ is the Higgs field vacuum expectation value.%
\footnote{In addition to the tree level contributions in Eq.~(\ref{eq:residual}), EFT loop corrections computed from the Feynman rules of Eq.~(\ref{EFT}) contribute to the physical mass splitting after electroweak symmetry breaking.  For renormalizable UV completions, the total correction at one loop order is~\cite{Cheng:1998hc,Feng:1999fu} $\Delta M = (-\alpha_2 M/\pi)[ [T^2 - (T^3)^2] f(m_W/M) + \frac{1}{c_W^2}(T^3 - s_W^2 Q)^2 f(m_Z/M) + s_W^2 Q^2 f(m_\gamma/M) ]$, where $T^a$ and $Y=Q-T^3$ are SU$(2)$ and U$(1)$ representation matrices, and $f(r)=\int_0^1 dx\,(2+2x)\log[x^2 + (1-x)r^2]$.  For the pure Higgsino case, loop corrections at order $1/M$ can be shown to vanish and the term $-av^2/(2M)$ yields the complete contribution to the mass splitting between $Q=1$ versus $Q=0$ states at this order.  }

Matching the UV theory (\ref{UV}) to the effective theory, similar to Ref.~\cite{Chen:2018uqz}, we obtain 
\ba
a&=&\frac{3}{2}\alpha_2^2 \frac{s_W^2}{c_W^2}+(\kappa_1^2+\kappa_2^2)\frac{M}{M^\prime-M} \,, \nonumber \\
b&=&(-\kappa_1^2+\kappa_2^2+2i\kappa_1\kappa_2)\frac{M}{M^\prime-M} \,, \nonumber \\
c&=&\frac{3}{4}\alpha_2^2\left(1-2\frac{s_W^2}{c_W^2}+\frac{1}{2 c_W^4}\right)-2(\kappa_1^2+\kappa_2^2)\frac{M}{M^\prime-M} \,,
\ea
where $\kappa_1$ and $\kappa_2$ are the Yukawa coupling constants in $F(H)$~\cite{Hill:2014yka},  $\alpha_2=g_2^2/(4\pi)$, $s_W=\sin{\theta_W}$
and $c_W=\cos{\theta_W}$ with $\theta_W$ the weak mixing angle. 
As we will see in the next section, the UV information relevant for low energy WIMP-matter scattering  is encoded in ($\kappa^2\equiv \kappa_1^2+\kappa_2^2$)
\begin{equation}
\tilde{c}_H \equiv
-(a+c-|b|)=-\frac{3}{4}\alpha_2^2 \left(1+\frac{1}{2c_W^4} \right)+ 2\kappa^2\frac{M}{M^\prime-M} \,.
\label{cH}
\end{equation}
If we had considered a heavy singlet instead of heavy triplet 
in the Standard Model extension (\ref{UV}), then $\tilde{c}_H$ would take an identical form to Eq.~(\ref{cH}), but with $M^\prime$ now representing
the singlet mass.%
\footnote{For the singlet extension, the non-gauge contributions are 
$\delta a = -(\kappa_1^2 + \kappa_2^2) M/(M^\prime - M)$, 
$\delta b = (-\kappa_1^2+\kappa_2^2 + 2i\kappa_1\kappa_2) M/(M^\prime -M)$
and 
$\delta c = 0$.
}

\section{Matching and RG evolution \label{sec:matching}}

In order to compute the cross section for dark matter direct detection at the nuclear level, we must match and evolve the electroweak scale effective theory of the WIMP specified in Eq.~(\ref{EFT}) to lower energy scales.  In a first step we integrate out weak scale particles $W^\pm$, $Z^0$, $h$ and $t$, and match to an effective theory consisting of five-flavor QCD, and the following effective interactions of the WIMP with quarks and gluons:%
\footnote{We restrict attention to elastic scattering.  Inelastic scattering~\cite{Bramante:2016rdh} could be investigated by considering operator structures $\bar{h}_0^{\rm high} h_0^{\rm low}$.}
\ba
\mathcal{L}=\bar{h}_0^{\rm low} h_0^{\rm low}\bigg \{\sum_{q=u,d,s,c,b}\big[c_q^{(0)}O_q^{(0)}+c_q^{(2)}v_{\mu}v_{\nu}O_q^{(2)\mu\nu} \big]+
c_g^{(0)}O_g^{(0)}+c_g^{(2)}v_{\mu}v_{\nu}O_g^{(2)\mu\nu} \bigg\} \,,
\ea
where the spin-0 and spin-2 quark and gluon operators are 
\begin{align}
  O^{(0)}_{q} &= m_{q} \bar{q}q \,, &
  O^{(2) \mu \nu }_{q} &= \frac{1}{2}\bar{q}\left(\gamma^{\{\mu}iD_{-}^{\nu\}}-\frac{g^{\mu \nu}}{d}i\slashed{D}_{-}\right)q \,,
  \nl
  O^{(0)}_{g} &= (G^{A}_{\mu\nu})^{2} \,, &
O^{(2) \mu \nu}_{g} &= -G^{A\mu\lambda}G^{A\nu}_{~~~\lambda}+\frac{1}{d}g^{\mu\nu}(G^{A}_{\alpha\beta})^{2} \,.
\label{eq:op}
\end{align}
We neglect operators of higher dimension that are suppressed by powers of 
hadronic scales times $1/m_W$ where $m_W$ is the mass of $W^\pm$ bosons.  Here $d=4-2\epsilon$  is the spacetime dimension,
$D_{-} \equiv \overrightarrow{D} - \overleftarrow{D}$,
and curly brackets around indices denote symmetrization.

\begin{figure}[t]
\centering
\includegraphics[width=.35\linewidth]{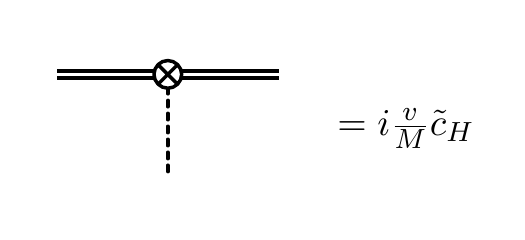}
\caption{ Feynman rule for 3-point interaction vertex involving the physical Higgs boson $h$ (dashed line) and the lightest electrically neutral Majorana fermion component of the Higgsino field, $h_0^{\rm low}$ (double line). The encircled cross denotes insertion of a $1/M$ effective theory vertex.
}
 \label{vertex}
\end{figure}

The matching process is similar to that in Ref.~\cite{Chen:2018uqz}, except that two boson $ZZ$ exchange contributes to the doublet case in addition to $WW$ exchange in the triplet case.  Also note that the WIMP-WIMP-Higgs boson three point vertex is the only contribution depending on UV parameter $\tilde{c}_H$, cf. Fig.~\ref{vertex}.%
\footnote{The coupling $\tilde{c}_H$ replaces $c_H$ for the triplet case in Ref.~\cite{Chen:2018uqz}.}
In terms of $\tilde{c}_H$, the renormalized matching coefficients in the $\overline{\rm MS}$ scheme for $n_f=5$ flavor QCD theory are:
\begin{align}\label{eq:results}
  \hat{c}_{U}^{(0)}(\mu)&= -\frac{1}{4 x_{h}^{2}} \left(1+\frac{1}{2c_W^3}\right) +\frac{1}{32 c_W}(c_V^{(U)2}-c_A^{(U)2})- \frac{m_W}{\pi M}\frac{\tilde{c}_H}{\alpha_{2}^{2} x_h^2}  \,,
  \nl
  \hat{c}_{D}^{(0)}(\mu)&= -\frac{1}{4 x_{h}^{2}} \left(1+\frac{1}{2c_W^3}\right) +\frac{1}{32 c_W}(c_V^{(D)2}-c_A^{(D)2})
  -\delta_{Db} \frac{x_{t}}{16(x_{t}+1)^{3}} 
  - \frac{m_W}{\pi M} \frac{\tilde{c}_H}{\alpha_{2}^{2} x_h^2}\,,
  \nl
 \hat{c}_{g}^{(0)}(\mu)&=    \frac{\alpha_s(\mu)}{4\pi}\bigg\{ \frac{1}{12}\bigg[\frac{1}{x_h^2}\Big(1+\frac{1}{2c_W^3}\Big)
  +1+\frac{1}{2(x_t+1)^2}\bigg] + \frac{m_W}{ \pi M}\frac{\tilde{c}_H}{3 \alpha_2^2 x_h^2} 
  +\frac{1}{64 c_W}\bigg[(c_V^{(D)2}+c_A^{(D)2}) \nl
&\quad +\frac{1}{4}(c_V^{(U)2}+c_A^{(U)2})\Big[\frac{8}{3}
+\frac{32y_t^6(8y_t^2-7)}{(4y_t^2-1)^{7/2}}\arctan(\sqrt{4y_t^2-1})
 -\pi y_t+\frac{4(48y_t^6-2y_t^4+9y_t^2-1)}{3(4y_t^2-1)^3}\Big] \nl
& \quad +\frac{1}{4}(c_V^{(U)2}-c_A^{(U)2})\Big[3\pi y_t -\frac{32y_t^4(24y_t^4-21y_t^2+5)}{(4y_t^2-1)^{7/2}}\arctan(\sqrt{4y_t^2-1}) \nl
& \quad -\frac{4(144y_t^6-70y_t^4+9y_t^2-2)}{3(4y_t^2-1)^3} \Big]\bigg] 
\bigg\}
  \,,
  \nl
  \hat{c}_{U}^{(2)}(\mu)&=\frac{1}{6} 
+\frac{1}{24 c_W}(c_V^{(U)2}+c_A^{(U)2})
 -\frac{1}{16 c_W^2}(c_V^{(U)2}+c_A^{(U)2})\frac{m_W}{\pi M}
-\frac{m_W}{4\pi M}  \,,
  \nl
  \hat{c}_{D}^{(2)}(\mu)&= \frac{1}{6} 
+\frac{1}{24  c_W}(c_V^{(D)2}+c_A^{(D)2})
 -\frac{1}{16  c_W^2}(c_V^{(D)2}+c_A^{(D)2})\frac{m_W}{\pi M}
-\frac{m_W}{4\pi M} \nl
&\quad +\frac{\delta_{Db}}{4}
\Big[ \frac{3x_t+2}{3(x_t+1)^3}-\frac{2}{3}-
\frac{m_W}{\pi M}\frac{x_t^2(1-x_t^4+4x_t^2\log{x_t})}{(x_t^2-1)^3} \Big]  \,,
  \nl
  \hat{c}_{g}^{(2)}(\mu)&=
  \frac{\alpha_s(\mu)}{4\pi}\bigg\{
 N_\ell\left( - {4\over 9} \log{\mu\over m_W} - {1\over 2} \right) - {(2+ 3x_t)\over 9(1+x_t)^3}\log{\mu \over m_W(1+x_t)}
\nl
&\quad
-{ ( 12 x_t^5 - 36 x_t^4 + 36 x_t^3 - 12 x_t^2 + 3 x_t - 2)\over 9 (x_t-1)^3}\log{x_t\over 1+x_t}
- {2 x_t ( -3 + 7 x_t^2) \over 9(x_t^2-1)^3} \log 2
\nl
&\quad
- { 48 x_t^6 + 24 x_t^5 - 104 x_t^4 - 35 x_t^3 + 20 x_t^2 + 13 x_t + 18 \over 36(x_t^2-1)^2 (1+x_t)}
  \nl
  &\quad
  + \frac{m_W}{4 \pi M}\bigg[
  N_\ell\left( \frac83 \log{\mu\over m_W} - \frac13 \right)
 + \frac{16 x_t^4}{3(x_t^2-1)^3} \log{x_t} \log{\mu\over m_W}
 - \frac{4(3x_t^2-1)}{3(x_t^2-1)^2}     \log{\mu \over m_W}
 + \frac{16 x_t^2}{3} \log^2{x_t}
 \nl
 &\quad
 - \frac{ 4( 4x_t^6-16x_t^4+6x_t^2+1)}{3(x_t^2-1)^3} \log{x_t}
 + \frac{ 8x_t^2(x_t^6-3x_t^4+4x_t^2-1)}{3(x_t^2-1)^3} {\rm Li}_2(1-x_t^2)
 + \frac{4\pi^2 x_t^2}{9}
 \nl
 &\quad
 - \frac{8 x_t^4 - 7 x_t^2 + 1}{3(x_t^2-1)^2} 
 \bigg] \nl
&\quad+ \frac{1}{64c_W}\bigg[\Big[2(c_V^{(U)2}+c_A^{(U)2})+3(c_V^{(D)2}+c_A^{(D)2})\Big]
\Big[-\frac{32}{9} \log{{\mu \over m_Z}} -4 \Big]\nl
&\quad+(c_V^{(U)2}+c_A^{(U)2})
\Big[\frac{32(24y_t^8-21y_t^6-4y_t^4+5y_t^2-1)}{9(4y_t^2-1)^{7/2}}\arctan(\sqrt{4y_t^2-1})-\frac{\pi y_t}{3}\nl
&\quad+\frac{4(48 y_t^6+62 y_t^4-47 y_t^2+9)}{9(4y_t^2-1)^3}\Big]
+(c_V^{(U)2}-c_A^{(U)2})\Big[\frac{4y_t^2(624y_t^4-538y_t^2+103)}{9(4y_t^2-1)^3}-\frac{13\pi y_t}{3}\nl
&\quad+\frac{32y_t^2(104y_t^6-91y_t^4+35y_t^2-5)}{3(4y_t^2-1)^{7/2}}\arctan(\sqrt{4y_t^2-1})\Big] \bigg] \nl
&\quad 
+\frac{1}{96 c_W^2} \frac{m_W}{\pi M}
\bigg[\Big[2(c_V^{(U)2}+c_A^{(U)2})+3(c_V^{(D)2}+c_A^{(D)2})\Big]\Big(8\log{{\mu \over m_Z}}-1\Big) \nl
&\quad -(c_V^{(U)2}+c_A^{(U)2})\Big[ 
\frac{1-18y_t^2+36y_t^4}{(4y_t^2-1)^2}+\frac{8(1-4y_t^2+3y_t^4+18y_t^6)\log{y_t}}{(4y_t^2-1)^3} \nl
&\quad +\frac{16 y_t^2(2-13y_t^2+32 y_t^4 -18y_t^6)}{(4y_t^2-1)^{7/2}}
\big[2 \arctan{\Big( \frac{1}{\sqrt{4y_t^2-1}} \Big)} \log{y_t} - {\rm Im}{{\rm Li}_2\Big(\frac{1-i\sqrt{4y_t^2-1}}{2 y_t^2}\Big)} \big] \Big]\nl
&\quad +4 y_t^2 (c_V^{(U)2}-c_A^{(U)2}) \Big[-\frac{8-59y_t^2+108 y_t^4}{(4y_t^2-1)^3} 
-\frac{(29-128y_t^2+108y_t^4)\log{y_t}}{(4y_t^2-1)^3} \nl
&\quad+ \frac{2(-7+38 y_t^2-82 y_t^4+108 y_t^6)}{(4y_t^2-1)^{7/2}}
\big[2 \arctan{\Big( \frac{1}{\sqrt{4y_t^2-1}} \Big)} \log{y_t} -{\rm Im}{{\rm Li}_2\Big(\frac{1-i\sqrt{4y_t^2-1}}{2 y_t^2}\Big)} \big]\Big]\bigg]
\Bigg\}
 \,. 
\end{align}
Here 
$c_V^{(U)}=1-\frac{8}{3}s_W^2$,
$c_V^{(D)}=-1+\frac{4}{3}s_W^2$,
$c_A^{(U)}=-1$,
$c_A^{(D)}=1$ with $U$ denoting up-type quarks, $D$ denoting down-type quarks.
The reduced coefficients $\hat{c}_i^{(S)}$ are given in terms of the original Wilson coefficients by $c_i^{(S)}\equiv (\pi\alpha_2^2/m_W^3) \hat{c}^{(S)}_i$, where $i = u,d,s,c,b,g$ is the index for quark or gluon and $S=0,2$ denotes the operator spin structure. The strong coupling is denoted by $\alpha_s(\mu)$. The mass ratios are defined as $x_j \equiv m_j/m_W$ and $y_j \equiv m_j/m_Z$ where $m_Z$ is the mass of $Z^0$ boson, and $j$ is the index of the specific particle, e.g. $j=t$ stands for top quark, $j=h$ for Higgs boson.
${\rm Li}_{2}(z) \equiv \sum_{k = 1}^{\infty} {z^k}/{k^2}$ is the 
dilogarithm function.
$N_\ell=2$ is the number of massless Standard Model generations. 
For electroweak scale matching, light quarks $u,d,s,c,b$ are treated as 
massless.  Also neglecting small corrections from $|V_{td}|^2$ and $|V_{ts}|^2$,
$u$ and $c$ quarks have the same coefficients, as do $d$ and $s$ quarks.  
We note that our results obey the correct formal limit at 
small $x_t$.%
\footnote{See Eq.~(9) of Ref.~\cite{Chen:2018uqz}.} 

For the evaluation of nucleon and nuclear matrix elements, the $n_f=5$ theory renormalized at the electroweak scale, $\mu_t \sim m_Z$ must be matched to $n_f=3$ renormalized at $\mu_0 = 1-2\,{\rm GeV}$.  We perform the necessary renormalization group evolution and threshold matching at heavy bottom and charm quark mass scales $\mu_b \sim m_b$ and $\mu_c \sim m_c$.  Details can be found in Ref.~\cite{Hill:2014yxa}.  Specifically, we take 
$\mu_{t} = (m_t+m_W)/2 = 126\,{\rm GeV}$, 
$\mu_{b} = 4.75\,{\rm GeV}$, 
$\mu_{c} = 1.4\,{\rm GeV}$, and 
$\mu_{0} = 1.2\,{\rm GeV}$, 
as in Refs.~\cite{Hill:2014yxa,Chen:2018uqz}. 
For the spin-0 coefficients, RG evolution and threshold matching are performed at NNNLO.
For spin-2 coefficients, the running and matching are at NLO.
We thus obtain a set of Wilson coefficients $c_i^{(S)}(\mu_0)$, for each of $u, d, s$ quarks and gluons, and for spin $S=0$ and $S=2$ operators. 

\section{Nucleon level cross section \label{sec:results}}

\begin{figure}
\centering
\includegraphics[width=.9\linewidth]{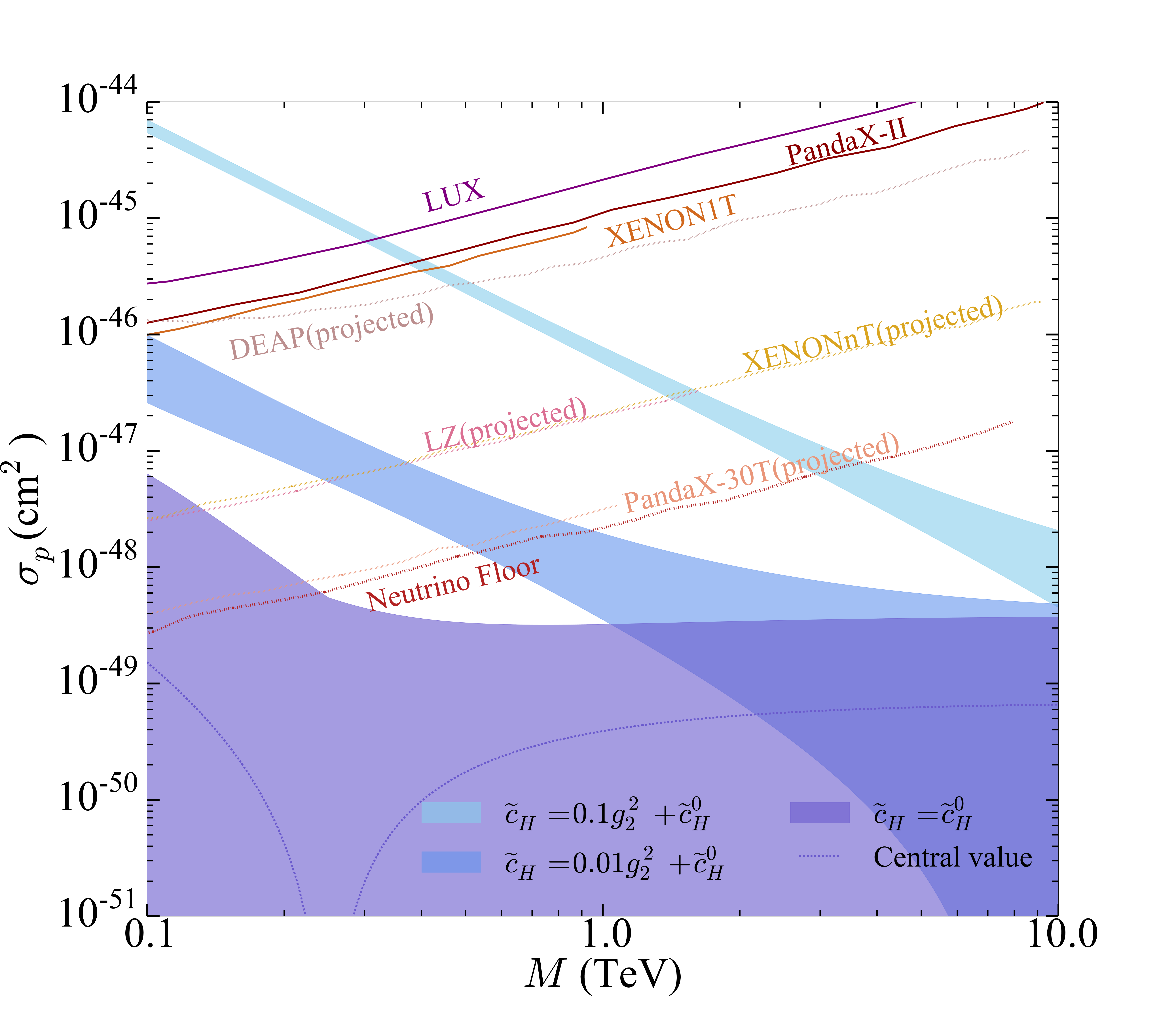}
\caption{
Scattering cross section for Majorana ${\rm SU}(2)$ doublet 
(Higgsino-like) WIMP on proton.
Corrections to this limit are parameterized by dimensionless Higgs coupling 
$\tilde{c}_H$ as discussed in the text.  
The pure Higgsino limit 
($\tilde{c}_H = -(3\alpha_2^2/4)[1+1/(2c_W^4)] \equiv \tilde{c}_H^0$)
is shown as the lower violet band and dashed central value curve.  
The impact of non-decoupled states in the UV completion are illustrated with 
$\tilde{c}_H= \tilde{c}_H^0 +  0.01\, g_2^2$ (middle, dark blue band) and  $\tilde{c}_H= \tilde{c}_H^0 + 0.1\, g_2^2 $ (upper, light blue band).  
Also shown are current dark matter direct detection experimental exclusion limits (solid lines)~\cite{Akerib:2016vxi,Cui:2017nnn, Aprile:2018dbl}, future projected detectors' sensitivities (dotted lines)~\cite{Aprile:2015uzo,Akerib:2015cja,Amaudruz:2017ekt, Liu:2017drf}, and neutrino floor (dash-dotted line) for Xenon detectors~\cite{Billard:2013qya}.
}
\label{rkappa}
\end{figure}

Having determined the low-scale quark level effective theory, let us proceed to evaluate hadronic matrix elements.  We begin by computing the low-velocity limit of the WIMP-nucleon scattering cross section, as a standard benchmark process.  Recall that 
\begin{equation}
\sigma_N  = \frac{m_{r}^{2}}{\pi}|\mathcal{M}^{(0)}_{N} + \mathcal{M}^{(2)}_{N}|^{2} \,,  
\end{equation}
where $N=n,p$ is a nucleon, $m_r = m_N M /(m_N + M) \approx m_N$ is the reduced mass of
the WIMP-nucleon system, and the scattering amplitude is
\begin{equation}
\mathcal{M}^{(S)}_{N} = \sum_{i= q,g} c^{(S)}_{i}(\mu_{0}) \langle N | O^{(S)}_{i}(\mu_{0})|N\rangle \,.
\end{equation}

Proton and neutron cross sections are identical, neglecting small corrections to isospin symmetry~\cite{Hill:2014yxa}.  
Lattice results~\cite{Aoki:2019cca} averaged from Refs.~\cite{Durr:2011mp,Yang:2015uis,Durr:2015dna, Freeman:2012ry, Junnarkar:2013ac} are taken for the light quark scalar matrix elements. 
Integrated parton distribution functions~\cite{Martin:2009iq} are taken for spin-2 matrix elements as in Ref.~\cite{Hill:2014yxa}.  
Quark masses are taken from lattice average results \cite{Aoki:2019cca}.
Explicitly, we have taken $\Sigma_{\pi N}=40(4)$~MeV, $R_{ud}=0.49(2)$ and $m_N f^{(0)}_{s,N}=56(8)~{\rm MeV}$.
In each case, small shifts are within the error bars assumed in Ref.~\cite{Hill:2014yxa}.%
\footnote{There is tension between the lattice result~\cite{Aoki:2019cca} 
$\Sigma_{\pi N}=40(4)$~MeV, and phenomenological extractions $\Sigma_{\pi N}=59.1(3.5)$~MeV and $\Sigma_{\pi N}=58(5)$~MeV from pionic atoms~\cite{Hoferichter:2015dsa} and  
low energy pion-nucleon scattering~\cite{RuizdeElvira:2017stg}, see also Ref.~\cite{Alarcon:2011zs}, which found $\Sigma_{\pi N}=59(7)$~MeV.  
Since the scalar matrix elements of $u$ and $d$ quarks represent a small
contribution to the total cross section~\cite{Hill:2011be}, 
our results would be essentially 
unchanged if a larger value, $\Sigma_{\pi N} \approx 60\,{\rm MeV}$, 
were used. }

For our default matching scales $\mu_t$, $\mu_b$, $\mu_c$ and $\mu_0$, and with the central values of all nucleon matrix elements at scale $\mu_0$, we find the spin-0 and spin-2 amplitudes for WIMP-proton scattering are (normalized by 
spin-2 amplitude  ${\cal M}^{(2)}_p|_{M\to\infty} = 1$)
\begin{align}\label{eq:amp_num}
  {\cal M}_p^{(0)} = -1.05 - 0.50 {\tilde{c}_H \over \alpha_2^2} {m_W \over M} \,, \qquad
  {\cal M}_p^{(2)} =  1 - 0.54  {m_W\over M} \,.
\end{align}
At $M\to\infty$, Eq.~(\ref{eq:amp_num}) exhibits a remarkable cancellation at the level of $\sim 5\%$ (compared to $\sim 20\%$ in the triplet case~\cite{Chen:2018uqz}).
At order $1/M$, the pure Wino case also exhibited a strong cancellation between the power corrections of spin-0 and spin-2 amplitudes~\cite{Chen:2018uqz}.   For the pure Higgsino case we have 
$\tilde{c}_H = -(3\alpha_2^2/4) [1+ 1/( 2 c_W^4)]$, and ${\cal M}_p^{(0)} = -1.05 +0.69\, {m_W/M}$.  Owing to the severity of the leading order cancellation, 
the total $1/M$ correction can compete with the leading order for moderate $M$.  
These features can be seen in the central value curve of Fig.~\ref{rkappa}, 
The sign of the power correction relative to leading power further suppresses the cross section as $M$ decreases from the heavy WIMP limit. 

As usual when evaluating the nucleon-level amplitude, we have two sources of uncertainties: Wilson coefficients and hadronic matrix elements.
Perturbative uncertainty in the matching coefficients is estimated by 
varying the matching scales within the ranges $m_W^2/2 \le \mu_t^2 \le 2 m_t^2$,
$m_b^2/2 \le \mu_b^2 \le 2 m_b^2$,
$m_c^2/2 \le \mu_c^2 \le 2 m_c^2$,
and
$1.0\,{\rm GeV} \le \mu_0 \le 1.4\,{\rm GeV}$, as in Ref.~\cite{Hill:2014yxa, Chen:2018uqz}. 
Uncertainties from neglected $1/M^2$ and higher order power corrections are estimated
by shifting ${\cal M}^{(2)}_p \to {\cal M}^{(2)}_p|_{M\to\infty}[ 1 \pm (m_W/M)^2 ]$
as in Refs.~\cite{Hill:2014yxa, Chen:2018uqz}.
Uncertainties from nucleon matrix elements are propagated to the observable cross section~\cite{Gasser:1982ap, Durr:2011mp, Junnarkar:2013ac, Martin:2009iq, Hill:2014yxa}. 
We add in quadrature the errors from different sources for each of the spin-0 and spin-2 amplitudes.   Then the maximum and minimum of all possible values of the combination $|\mathcal{M}^{(0)}_{p} + \mathcal{M}^{(2)}_{p}|$ sets the boundaries of the cross section curves depicted in Fig.~\ref{rkappa}.%
\footnote{We combined errors in this way at the amplitude level since 
the cross section can be zero for some parameter values.}

For the lower, violet colored, region in Fig.~\ref{rkappa}, the limit $M^\prime \gg M$ has been taken to decouple heavier states in the pure Higgsino limit for the coefficient $\tilde{c}_H$ in Eq.~(\ref{eq:amp_num}).  Before taking this limit, we may use Eq.~(\ref{cH}) to investigate the impact of non-decoupled states in the UV completion.  
Away from the pure-state limit, we have 
$\tilde{c}_H \approx 2\kappa^2 {M}/({M^\prime-M})$. 
For weakly coupled theories we consider a range of values 
$\tilde{c}_H / g_2^2 = 0.01 - 0.1$
in Fig.~\ref{rkappa}.  
For TeV mass WIMPs, the cross section is within the detectable range of next generation detectors~\cite{Aprile:2015uzo, Akerib:2015cja, Liu:2017drf} when $\tilde{c}_H/g_2^2 \approx 0.1$, and is close to the neutrino floor when 
$\tilde{c}_H/g_2^2 \approx 0.01$. 
In the pure Higgsino limit, the cross section upper limit remains at or below  $10^{-48}\,\rm{cm^2}$ for masses above a few hundred GeV.  

\section{Nuclear effects}
\label{sec:nuclear}
\begin{figure}
\centering
\includegraphics[width=.49\linewidth]{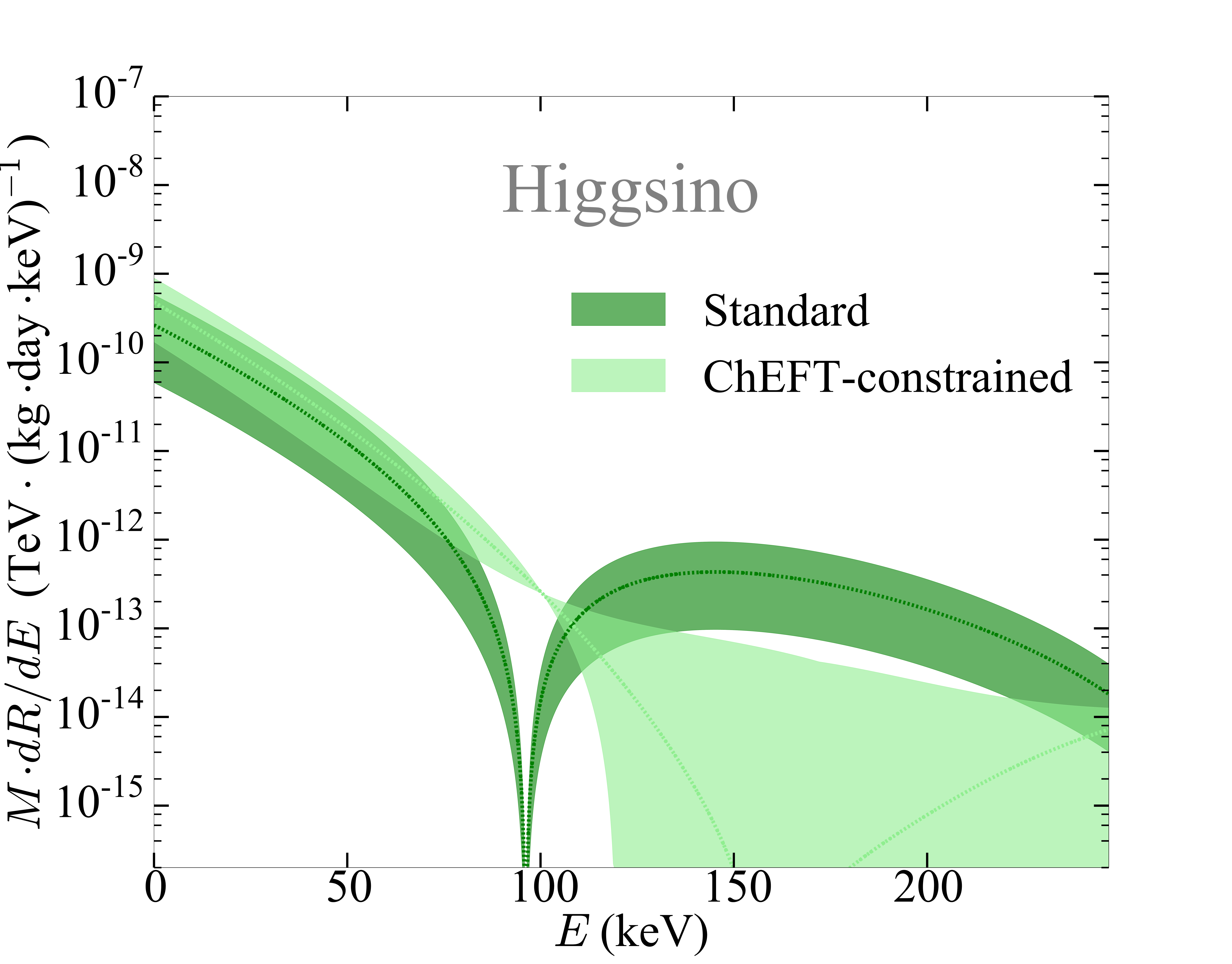}
\hspace{0.1cm}
\centering
\includegraphics[width=.49\linewidth]{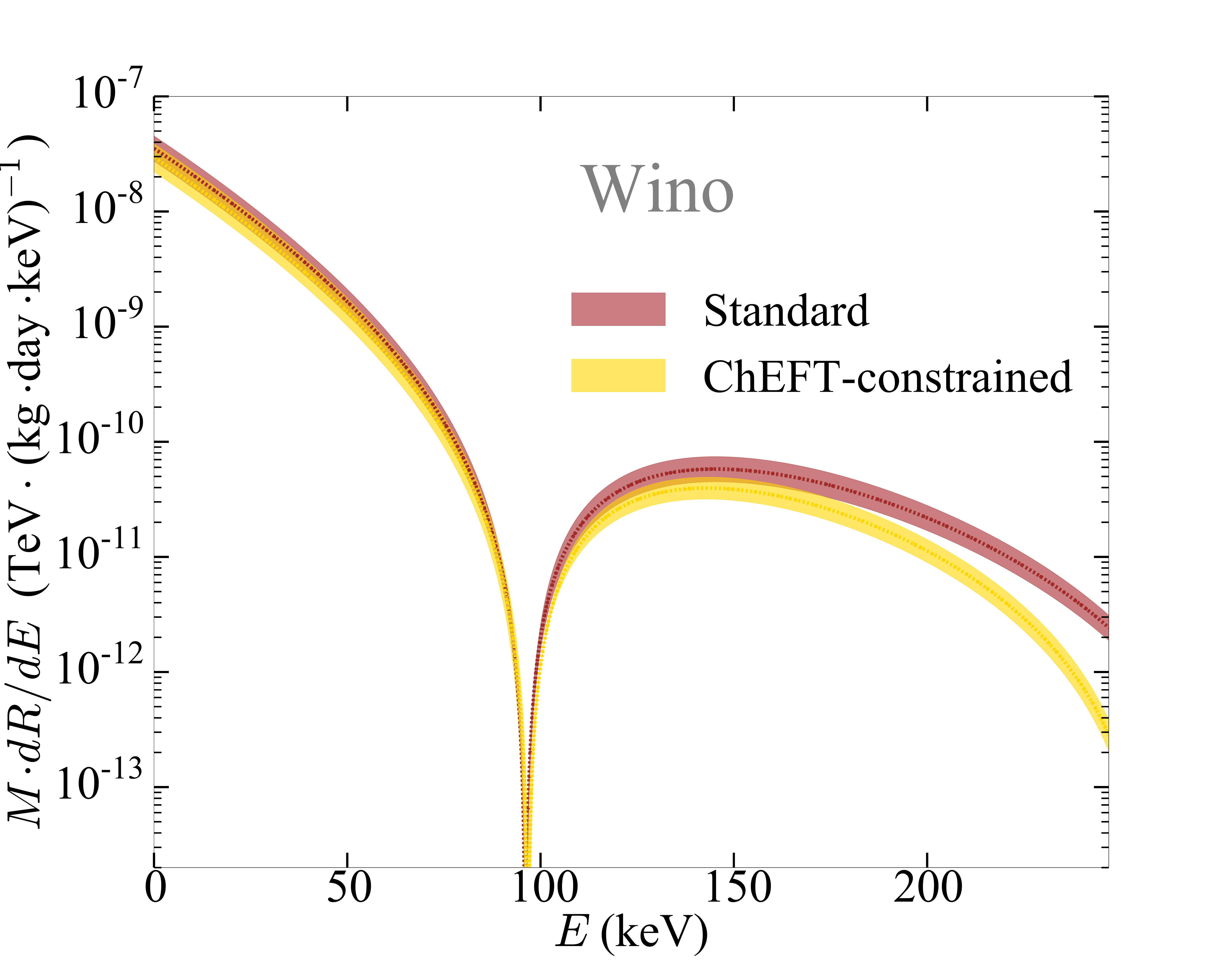}
\caption{Event rate (times WIMP mass) versus recoil energy for Xenon target, 
for Higgsino-like (left panel) and Wino-like (right panel) WIMP.
The ``Standard" rate uses the standard halo model, dark matter velocity distribution and Helm nuclear form factor~\cite{Lewin:1995rx}. The ``ChEFT constrained" rate replaces the Helm form factor by the model of Refs.~\cite{Hoferichter:2016nvd,Hoferichter:2018acd}.
Dashed curves are central values and shaded regions represent perturbative matching uncertainty.  Nucleon matrix element uncertainties are not displayed. 
}
\label{rate}
\end{figure}

\begin{figure}
\centering
\includegraphics[width=.49\linewidth]{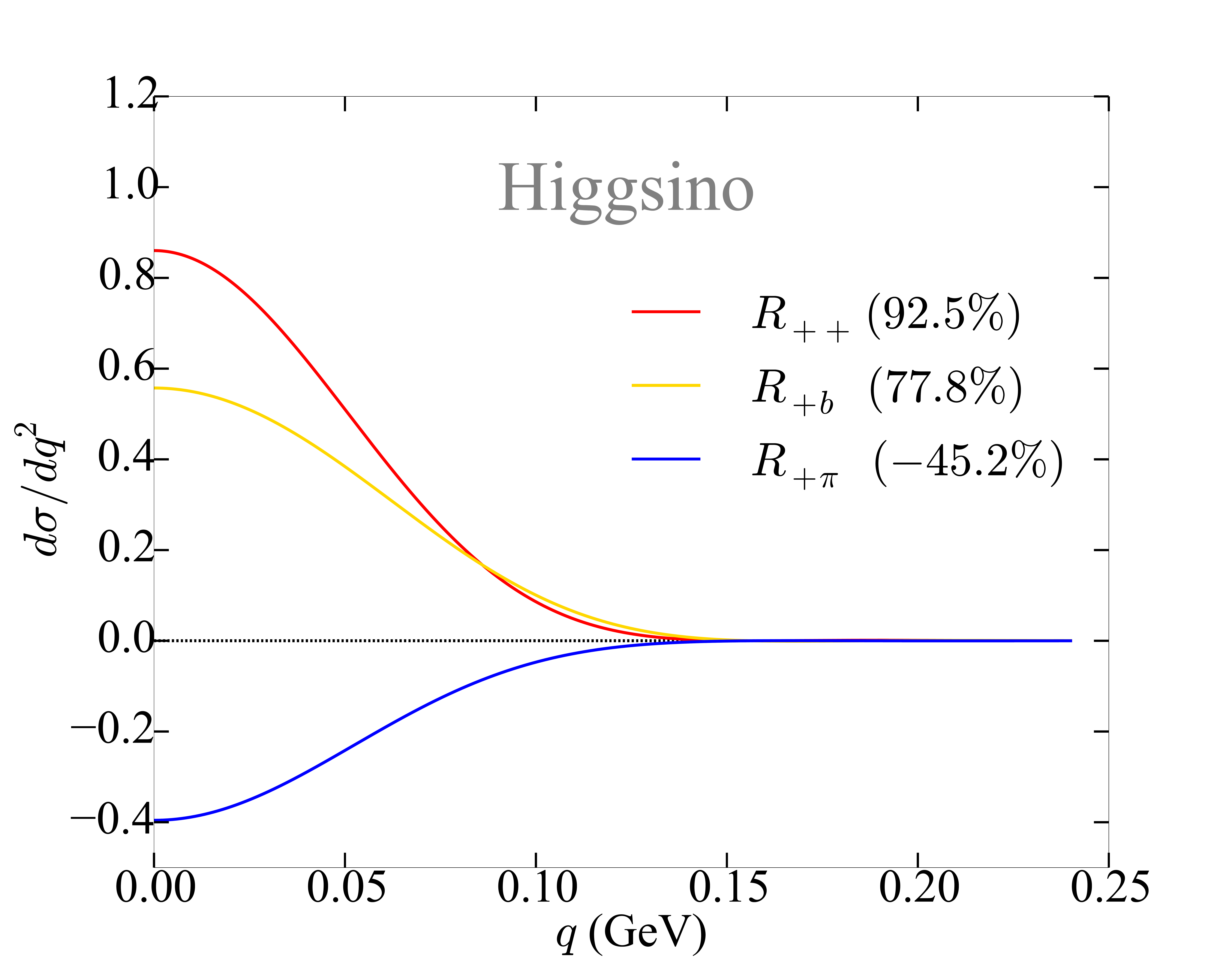}
\hspace{0.1cm}
\centering
\includegraphics[width=.49\linewidth]{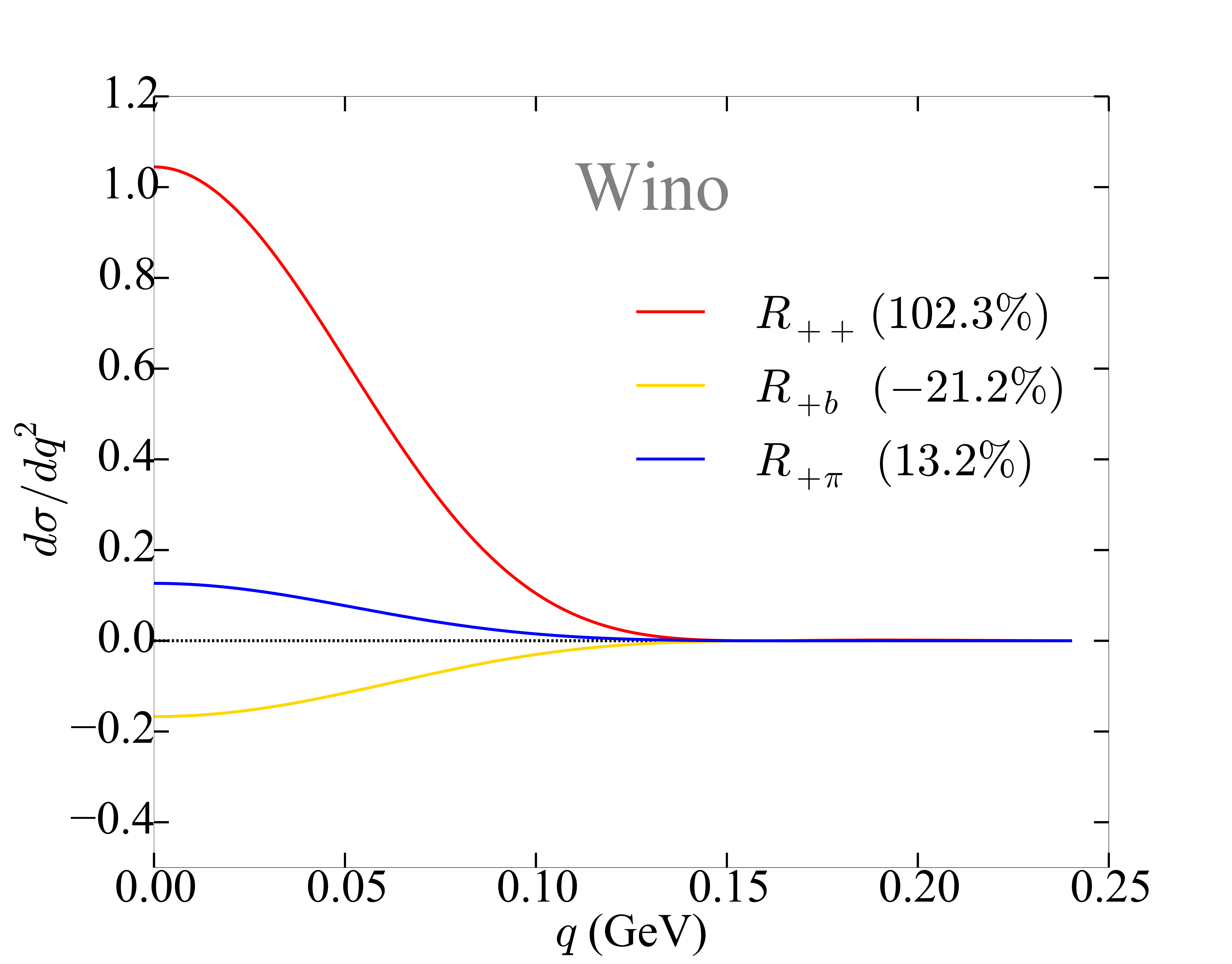}
\caption{Different nuclear response channels contributing to $d \sigma /d q^2$ versus momentum transfer $q$, for Higgsino-like (left panel) and Wino-like (right panel) WIMPs of velocity $v/c=10^{-3}$ scattering on a Xenon nucleus.  Curves are normalized to unity for the sum of all contributions at $q=0$. 
$R_{++}$ denotes the one-body isoscalar response 
(denoted $(c^M_+ \mathcal{F}_+^M)^2$ in Ref.~\cite{Hoferichter:2018acd}).
$R_{+b}$ and $R_{+\pi}$ denote one-body/two-body interference terms
(denoted $2\,c^M_+ c_b \mathcal{F}_+^M\mathcal{F}_b$ and $2\,c^M_+ c_\pi \mathcal{F}_+^M\mathcal{F}_\pi$ in Ref.~\cite{Hoferichter:2018acd}).
Numbers in parentheses are the percentage contribution for each nuclear response channel to the total integrated cross section.  
Other subdominant channels are not displayed. 
}
\label{channel}
\end{figure}

Finally, we note that the actual scattering process in experimental searches 
takes place on compound nuclei versus on isolated nucleons.  A standard practice for treating nuclear modifications is to apply a nuclear form factor to the free-nucleon result~\cite{Helm:1956zz, Lewin:1995rx}.  At the same time, a distribution in WIMP velocities is assumed present in our local galactic halo, and the event rate ansatz is a convolution of halo velocity profile, nuclear form factor, and single nucleon cross section~\cite{Lewin:1995rx}.

The severe cancellation in the single-nucleon cross section, 
cf. Eq.~(\ref{eq:amp_num}), suggests that nuclear corrections could potentially 
have a larger than expected impact.  We investigate this possibility here, 
using the model presented in Refs.~\cite{Hoferichter:2018acd, Hoferichter:2016nvd} 
which includes effects from multi-nucleon interactions.
In general, the differential event rate versus recoil energy takes the form
\be{\frac{dR}{dE}=\frac{\rho}{2\pi M}
|\mathcal{F}(q^2)|^2\int_{v_{min}(E)}^{\infty}\frac{f(\bf{v})}{v}d^3v
}
\,,
\label{rateeq}
\ee
where $\rho$ is the dark matter mass density, $\mathcal{F}(q^2)$ is the nuclear structure factor~\cite{Hoferichter:2018acd, Hoferichter:2016nvd} and $f(\bf{v})$ is the WIMP velocity distribution function.
Equation~(\ref{rateeq}) reduces to the ``Standard" treatment under 
the replacement $|\mathcal{F}(q^2)|^2 \to A^2|F(q^2)|^2\pi\sigma_N M/m_r^2$, where $F(q^2)$ is the Helm form factor~\cite{Helm:1956zz} and $A$ is the mass number of the nucleus. 
Figure~\ref{rate} compares the event rate for Xenon detectors (${}^{132} \, \rm{Xe}$) from our ``Standard" computation, to the ``ChEFT constrained" computation with nuclear model from Refs.~~\cite{Hoferichter:2018acd, Hoferichter:2016nvd}.%
\footnote{For discussion of nuclear corrections see also Refs.~\cite{Cirigliano:2012pq,Korber:2017ery}.}
The rate is multiplied by the WIMP mass to make the curves on the plot independent of WIMP mass.%
\footnote{For the purpose of illustrating nuclear effects, we focus on the heavy WIMP limit, neglecting small corrections to both curves when $m_W/M \ne 0$.
}
In order to perform an ``apples to apples" comparison in Fig.~\ref{rate}, 
we have employed the same input values for nucleon-level matrix elements in our ``Standard" nuclear model as in Ref.~\cite{Hoferichter:2016nvd}, except for the spin-0 gluon matrix element, which was evaluated in Ref.~\cite{Hoferichter:2016nvd} using a leading order perturbative QCD 
relation.  Higher order perturbative QCD corrections turn out to be significant~\cite{Hill:2013hoa,Hisano:2015rsa}, and we have included  corrections through NNNLO in both ``Standard" and ``ChEFT-constrained" analyses. 

We have also implemented the ``ChEFT constrained" nuclear model for the triplet case studied in Ref.~\cite{Chen:2018uqz}.  
For both doublet and triplet cases, 
a breakdown of $d\sigma/dq^2$ into separate nuclear response channels is displayed in Fig.~\ref{channel}, for an illustrative WIMP velocity $v/c=10^{-3}$. 
In the terminology of Ref.~\cite{Hoferichter:2018acd}, isoscalar one-body scattering (corresponding to the ``Standard" nuclear form factor treatment) 
remains the dominant contribution to the direct detection rate.   
As the comparison of Higgsino-like and Wino-like cases in the figure illustrates, 
the more severe cancellation of the Higgsino one-body amplitude, Eq.~(\ref{eq:amp_num}), 
leads to an effective enhancement of one-body/two-body interference terms in the Higgsino-like 
cross section.%
\footnote{An interesting feature of these plots is a partial cancellation between the 
different two-body contributions, $R_{+b}$ and $R_{+\pi}$.  This cancellation is not directly
related to the cancellation between spin-0 and spin-2 one body amplitudes (\ref{eq:amp_num}).
The ratio of the relevant two-body nuclear response coefficients can be expressed in our basis as 
$c_\pi : c_b = 
[ \sum_{q=u,d} c_q^{(0)} f_q^\pi - 8\pi c_g^{(0)}/(9\alpha_s) -(1/2) \sum_{i=u,d,s,g} c_i^{(2)} f_{i,\pi}^{(2)} ]
: 
[ - 8\pi c_g^{(0)}/(9\alpha_s) + (1/4) \sum_{i=u,d,s,g} c_i^{(2)} f_{i,\pi}^{(2)}]$, 
where $f_q^\pi$ and $f_{q,\pi}^{(2)}$ are defined in Ref.~\cite{Hoferichter:2018acd}.  
Coefficients $c_i^{(0)}$ and $c_i^{(2)}$ 
enter with different weights in the two-body responses, compared 
to the one-body response which involves
$c_+^M \propto \frac12 \sum_{N=p,n} \sum_{i=u,d,s,g} 
[ c_i^{(0)} f_{i,N}^{(0)} + \frac34 c_i^{(2)} f_{i,N}^{(2)}]$, 
where $f_{i,N}^{(0)}$ and $f_{i,N}^{(2)}$ are defined 
in Ref.~\cite{Hill:2014yxa}. 
}

However, for both doublet and triplet cases, the considered nuclear modifications do not qualitatively impact the cross section, and are comparable in magnitude to other sources of uncertainty.

\section{Summary \label{sec:summary}}

We have computed subleading corrections to the cross sections for Wino-like and Higgsino-like WIMP particles scattering on nuclear targets. 
The doublet result shows that order $1/M$ corrections do not significantly 
enhance  the small leading order cross section. The upper limit for the pure doublet is less than $10^{-48}\, \rm{cm}^2$ in the TeV mass regime, consistent with the leading order estimate~\cite{Hill:2013hoa,Hill:2014yxa}, and lower than the estimated Xenon neutrino floor~\cite{Billard:2013qya}.
We also investigated the impact of nuclear effects by employing a modern EFT-based model in place of the standard nuclear form factor. Although individual nuclear response channels have significant contributions, cancellations occur in the total nuclear response.
The overall effect of the new nuclear model is comparable to other uncertainties for both Wino-like and Higgsino-like WIMP candidates, leaving the TeV scale pure Wino within striking distance of next generation experiments, and the pure Higgsino well below the neutrino floor. 
The small cross section in the pure Higgsino limit can be modified by non-decoupled states in the UV completion.  For WIMP masses of order $M =500\,{\rm GeV}$, current experiments constrain the dimensionless Higgs-WIMP-WIMP coupling as $\tilde{c}_H \lesssim 0.04$. 

\vskip 0.2in
\noindent
{\bf Acknowledgements.}
We thank M.~Hoferichter and M.~Solon for discussions and 
comments on the manuscript.  
QC also thanks Xinshuai Yan for helpful discussions.    
Work supported by the U.S. Department of Energy, Office of
Science, Office of High Energy Physics, under Award Number
DE-SC0019095. Fermilab is operated by Fermi Research Alliance, LLC
under Contract No. DE-AC02-07CH11359 with the United States Department
of Energy. QC thanks the Fermilab theory group, and the high energy theory group at Northwestern University, for hospitality during visits where part of this work was performed.
\vskip 0.1in
\noindent

\bibliography{higgsino_1overM}

\end{document}